\documentclass[12pt]{article}
\usepackage{epsfig}

\newcommand{\tr}{{\rm tr}}
\newcommand{\pfrac}[2]{\left(\frac{#1}{#2}\right)}

\newcommand{\GeV}{{\rm\,GeV}}

\newcommand{\ice}[1]{\relax}

\begin{document}
\thispagestyle{empty}
\begin{flushright}
MZ-TH/01-10\\
hep-ph/0103313\\
March 2001\\
\end{flushright}
\vspace{0.5cm}
\begin{center}
{\Large\bf Heavy quark induced effective action for\\[-5pt]
  gauge fields in the $SU(N_c)\otimes U(1)$ model and the\\[5pt]
  low-energy structure of heavy quark current correlators}\\[32pt]
{\large S.~Groote$^1$ and A.A.~Pivovarov$^{1,2}$}\\[12pt]
$^1$ Institut f\"ur Physik der Johannes-Gutenberg-Universit\"at,\\[3pt]
Staudinger Weg 7, 55099 Mainz, Germany\\[7pt]
$^2$ Institute for Nuclear Research of the\\[3pt]
Russian Academy of Sciences, Moscow 117312
\end{center}
\vspace{1cm}
\begin{abstract}\noindent
We calculate the low-energy limit of heavy quark current correlators within
an expansion in the inverse heavy quark mass. The induced low-energy currents
built from the gluon fields corresponding to the initial heavy quark currents 
are obtained from an effective action for gauge fields in the one-loop
approximation at the leading order of the $1/m$ expansion. Explicit formulae
for the low-energy spectra of electromagnetic and tensor heavy quark current
correlators are given. Consequences of the appearance of a nonvanishing
spectral density below the two-particle threshold for high precision
phenomenology of heavy quarks are discussed quantitatively.
\end{abstract}

\section{Introduction}
Quantum corrections can qualitatively change the analytic structure of the
Green function or the symmetry properties of a classical field-theoretical
system. Qualitatively new features compared to the tree level picture may
emerge already at finite orders of perturbation theory for Green functions
while some effects can only appear as a result of summing up an infinite
number of perturbative terms. An effect of non-conservation of the Abelian
axial current reveals itself at leading order as a result of the calculation
of an one-loop triangle diagram~\cite{aban,const} while spontaneous symmetry
breaking or bound state formation cannot be observed at any finite order of
perturbation theory and requires an infinite summation of relevant subsets of
diagrams. For problems related to an investigation of the symmetry properties
of quantum systems, such an infinite summation can be readily done by
introducing an effective action for the system and calculating it as a loop
expansion series. Such an approach reorders the perturbation series with
respect to the lines of diagrams related to external fields and allows one to
take into account the entire dependence on external fields exactly within a
given order of the loop expansion. An effective action can be considered as a
generating functional for the vertex (one-particle irreducible, or proper)
Green functions (see e.g.\ Ref.~\cite{bog}). The treatment of external fields
beyond finite order of expansion was done in Ref.~\cite{first}. The efficient
method of calculating the effective action is based on the Legendre transform
of the generating functional for connected Green functions and heavily uses
the functional techniques~\cite{funk}. One can also use a practical
calculation technique by substituting the shifted fields in the original
Lagrangian of the system~\cite{jack,tadp}. The part of the effective action
which is constructed from the constant field is usually called an effective
potential and is used to analyze the fundamental symmetry properties of the
theory beyond the plain perturbation theory where the effect of the external
fields is resummed to all orders. The expansion leading to the effective
potential is, in fact, an expansion in Planck's constant $\hbar$, i.e.\ the
correction accounts for the deflection from the classical limit. Therefore new
effects which are absent in the classical approximation may appear within such
an approach. An example for this kind of new quantum effects is the
light-by-light scattering which emerges as a quantum correction to the photon
dynamics due to the interaction with virtual electrons. In the low-energy
limit it can be seen as a nonlinearity of the equations for the strong
electromagnetic fields in the vacuum. The behaviour of the electromagnetic
fields with such a correction can be described by the Euler-Heisenberg
Lagrangian~\cite{Heisenberg}. The generalization to non-Abelian fields is
discussed in Ref.~\cite{savvidy}. The effective potential is also a powerful
tool for investigating effects of spontaneous symmetry breaking by quantum
corrections~\cite{coleweinb} and for analyzing the properties of particle
systems at finite temperature and density~\cite{linde}.

A special advantage of using the external field technique in gauge theories is
an explicit gauge invariance of the effective action that allows to
drastically simplify the computation and to reduce the number of necessary
diagrams~\cite{ext}. It is generally believed that in non-Abelian gauge
theories the nonperturbative fluctuations (instantons)~\cite{insta} create a
complex vacuum structure that eventually explain (or is responsible for) the
low-energy observable spectrum of particles~\cite{complvac}. The technique of
calculating in external fields was heavily used for the calculation of quantum
corrections to the effective action of gauge fields in the classical instanton
background~\cite{thooft}. In practical applications to QCD the complex vacuum
structure could explain a phase transition from the quark-gluon representation
of Green functions at high energies to the hadron picture at low energies.
While the problem of a full description of this transition remains unresolved,
Wilson's operator product expansion which is one of the key tools for
calculating the correlation functions at short distances is used for
describing hadronic properties at low energies in a semiphenomenological way
using sum rule techniques~\cite{wilson}. The external field technique provides
a convenient way for practical calculations within the sum rule method based
on a semiphenomenological account for the condensates of the local
operators~\cite{svz,fortsc}.

In the present paper we calculate the low-energy limit of heavy quark current
correlators within the $1/m$ expansion where $m$ is a heavy quark mass. The
induced low-energy currents corresponding to the heavy quark initial operators
are obtained from an effective action for gauge fields of the
$SU(N_c)\otimes U(1)$ model in the one-loop approximation at leading order of
both the coupling constant and the $1/m$ expansion. Our results for the
effective action for vector and tensor currents are presented in Sec.~2. Using
the external field technique as a convenient framework for practical
calculations of the induced currents in this model, in Sec.~3 we present these
induced currents. In Sec.~4 we show phenomenological applications, and in
Sec.~5 we deal with the consequences encountered for calculating arbitrary
moments.

\section{The effective action}
While the technique is standard, we are aiming at concrete results for further
phenomenological applications to sum rules for the vacuum polarization
functions of heavy quarks. Therefore we briefly outline the calculation and
the related issues. Further details can be found in Refs.~\cite{bog,Itzykson}. 
The Lagrangian of a heavy fermion field $\psi$ interacting with a gauge field
${\cal B}$ of the gauge group $SU(N_c)\otimes U(1)$ reads
\begin{equation}
L=\bar\psi\left(i\gamma^\mu\partial_\mu+\gamma^\mu{\cal B}_\mu-m\right)\psi
\end{equation}
where ${\cal B}_\mu=eA_\mu+g_sB_\mu$. Here $A_\mu$ is a gauge field of the
$U(1)$ subgroup (photon) with the coupling constant $e$ and $B_\mu$ is a gauge
field of the $SU(N_c)$ subgroup (gluon) with the coupling constant $g_s$. The
matrix notation for the non-Abelian gauge field potentials is used,
$B_\mu=t_aB^a_\mu$, $t_a$ are generators of the gauge group $SU(N_c)$. A
generating functional $W[J]$ of connected Green functions is given by a
functional integral with the sources $J$,
\begin{eqnarray}
Z[J]&=&\exp(iW[J])\ =\ \int[d\bar\psi\,d\psi]\exp\left(i\int L^J(x)d^4x\right)
  \nonumber\\&=&\int[d\bar\psi\,d\psi]\exp\left(i\int\left(\bar\psi(i\gamma^\mu
  \partial_\mu+\gamma^\mu{\cal B}_\mu-m)\psi
  +J^\mu{\cal B}_\mu\right)d^4x\right)
\end{eqnarray}
where the product $J^\mu{\cal B}_\mu$ implies a trace with respect to the
representation of $SU(N_c)\otimes U(1)$. A proper gauge fixing is implied as
well. The effective action $\Gamma[\bar{\cal B}]$ for the gauge field is then
given by the Legendre transform
\begin{equation}
\Gamma[\bar{\cal B}]=W[J]-J\bar{\cal B},\qquad
  \bar{\cal B}=\frac{\delta W[J]}{\delta J}.
\end{equation}
It was shown that this procedure is equivalent to the more direct calculation
in external fields~(see e.g.\ Ref.~\cite{Itzykson}). It is also a
generalization of results obtained for constant external fields~\cite{const}.
Up to leading order in $\hbar$ the effective action constructed with a
Legendre transform can also be found through
\begin{equation}
\exp(i\Gamma[{\cal B}])=\int[d\bar\psi\,d\psi]\exp\left(i\int\bar\psi
  (i\gamma^\mu\partial_\mu+\gamma^\mu{\cal B}_\mu-m)\psi\,d^4x\right)
  =\det\left(i\gamma^\mu\partial_\mu+\gamma^\mu{\cal B}_\mu-m\right)
\end{equation}
where ${\cal B}$ is now a classical gauge field (integration over $d^4x$ is
implied for expressions of the action). By using the identity
$\det M=\exp(\tr(\ln M))$ for an operator $M$ one continues with
\begin{equation}
i\Gamma[{\cal B}]=\tr\Big[\ln\left
  (i\gamma^\mu\partial_\mu+\gamma^\mu{\cal B}_\mu-m\right)\Big].
\end{equation}

\subsection{Results for the effective action}
\begin{figure}\begin{center}
\epsfig{figure=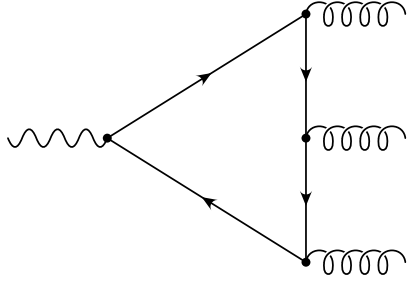, scale=0.4}
\caption{\label{fig1}Heavy quark loop correction to the electromagnetic
  current}\end{center}
\end{figure}

A straightforward calculation of the functional determinant gives a correction
to the effective low-energy action of the gauge fields within
$SU(N_c)\otimes U(1)$. We calculate the leading nontrivial contribution in the
$1/m$ expansion. The functional determinant in loop expansion can be
represented by Feynman diagrams. A diagram which gives a correction to the
effective action due to a heavy quark loop is shown in Fig.~\ref{fig1}.
Two-gluon transitions are forbidden according to a generalization of Farry's
theorem to non-Abelian theories~\cite{Farri}. We are interested in the
behaviour of the amplitude associated with the diagram in Fig.~\ref{fig1} at
low energies and, therefore, take the limit of a very heavy quark. Formally
the limit $m\to\infty$ is taken which in physical terms means that $m$ is much
larger than all momenta of external legs of the diagram in Fig.~\ref{fig1},
namely the three gluons and the photon. A straightforward calculation of the
diagram presented in Fig.~\ref{fig1} gives the one-loop result for the
correction to the effective action induced by a heavy quark loop at leading
order of the $1/m$ expansion. It reads
\begin{equation}\label{actqcd}
\Delta\Gamma_{\rm QCD}=\frac{eg_s^3d_{abc}}{180m^4(4\pi)^2}
  \Big[14\tr(FG^aG^bG^c)-5\tr(FG^a)\tr(G^bG^c)\Big]
\end{equation}
where $F_{\mu\nu}$ is a field strength tensor for the $U(1)$ subgroup,
$G_{\mu\nu}=t_aG^a_{\mu\nu}$ is the field strength tensor for the $SU(N_c)$
subgroup, and $d_{abc}$ are the totally symmetric $SU(N_c)$ structure
constants defined by the relation $d_{abc}=2\tr(\{t_a,t_b\}t_c)$. The trace in
Eq.~(\ref{actqcd}) is understood as a trace with respect to the Lorentz indices
of the fields, i.e.\ one considers the strength tensors of gauge fields as
matrices for which $\tr(FG^a)=F^{\mu\nu}G^a_{\nu\mu}$. This makes the formulae
shorter and more transparent. Before we proceed to the calculation of the
induced current in the next section, we compare the result with the
corresponding expression in QED.

\subsection{Comparison with QED}
The effective action within QED corresponding to Eq.~(\ref{actqcd}) is known
as Euler-Heisenberg Lagrangian~\cite{Itzykson},
\begin{equation}\label{actqed}
\Delta\Gamma_{\rm QED}=\frac{2\alpha^2}{45m^4}\Big[(\vec E^2-\vec H^2)^2
  +7(\vec E\cdot\vec H)^2\Big],\qquad\alpha=\frac{e^2}{4\pi}
\end{equation}
which can be obtained from Eq.~(\ref{actqcd}) by replacing $d_{abc}$ by $1$
and $G^a_{\mu\nu}$ by $F_{\mu\nu}$, supplemented by the obvious substitution
$g_s\to e$. Using the definitions for the electric field $\vec E$ and the
magnetic field $\vec H$,
\begin{equation}
F^{0j}=-F_{0j}=-E_j,\quad
F^{i0}=-F_{i0}=E_i,\quad
F^{ij}=F_{ij}=-\epsilon_{ijk}H_k
\end{equation}
one finds
\begin{eqnarray}
\tr(F^2)=2(\vec E^2-\vec H^2),&&
\tr(F^4)=2(\vec E^2-\vec H^2)^2+4(\vec E\cdot\vec H)^2,\nonumber\\
\tr(F\tilde F)=4(\vec E\cdot\vec H),&&
\tr(F\tilde FF\tilde F)=-\frac12(\tr(F^2))^2+\tr(F^4)=4(\vec E\cdot\vec H)^2
\end{eqnarray}
where
\begin{equation}
\tilde F_{\mu\nu}=\frac12\epsilon_{\mu\nu\rho\sigma}F^{\rho\sigma}.
\end{equation}
Note for further use and convenience that in order to rewrite the expressions
from one form to another there is one more relation between the fourth-order
monomials of the electromagnetic field strength tensor $F_{\mu\nu}$,
\begin{equation}
(F\tilde F)^2=-2(F^2)^2+4F^4
\end{equation}
with 
\begin{equation}
F\tilde F=F^{\mu\nu}\tilde F_{\mu\nu},\qquad
F^2=F^{\mu\nu}F_{\mu\nu},\qquad
F^4=F^{\mu\nu}F_{\nu\alpha}F^{\alpha\beta}F_{\beta\mu}.
\end{equation}
Therefore the correspondence between Eq.~(\ref{actqcd}) and its QED
counterpart in Eq.~(\ref{actqed}) is established.

\section{The induced current}
An expression for the induced electromagnetic current $J^\mu$ as being an
effective electromagnetic current for the low-energy effective theory
describing the interaction of photons and gluons is given by the derivative of
the effective action with respect to the external Abelian gauge field,
\begin{equation}
e J^\mu=-\frac{\delta\Gamma[{\cal B}]}{\delta A_\mu}
=-\tr\left[ie\gamma_\mu\,\frac1{i\gamma^{\mu'}\partial_{\mu'}
  +\gamma^{\mu'}{\cal B}_{\mu'}-m}\right].
\end{equation}
The derivative with respect to $A_\mu$ can be replaced by a derivative with
respect to $F_{\mu\nu}=\partial_\mu A_\nu-\partial_\nu A_\mu$, 
\begin{equation}
eJ^\mu(x)=-\frac{\delta F_{\mu'\nu'}}{\delta A_\mu}\
  \frac{\delta\Gamma[{\cal B}]}{\delta F_{\mu'\nu'}}
  =-2\partial_\nu\frac{\delta\Gamma[{\cal B}]}{\delta F_{\nu\mu}}.
\end{equation}

\subsection{Results for the induced vector current}
With the explicit expression for the effective action given in
Eq.~(\ref{actqcd}) we obtain~\cite{grpiv}
\begin{equation}\label{indvec}
J^\mu=\partial_\nu{\cal O}^{\mu\nu},\qquad
{\cal O}^{\mu\nu}=\frac{-g_s^3d_{abc}}{90m^4(4\pi)^2}[14(G^aG^bG^c)^{\mu\nu}
  -5(G^a)^{\mu\nu}\tr(G^bG^c)].
\end{equation}
Note that the current conservation $\partial_\nu J^{\mu\nu}=0$ is
automatically guaranteed because the operator ${\cal O}^{\mu\nu}$ is
antisymmetric, ${\cal O}^{\mu\nu}+{\cal O}^{\nu\mu}=0$. Higher order
corrections in the coupling constant $\alpha$ of the $U(1)$ subgroup are
omitted. The induced electromagnetic current in Eq.~(\ref{indvec}) is a
correction of order $1/m^4$ in the inverse heavy quark mass which vanishes
in the limit of an infinitely heavy quark. Corrections in the inverse heavy
quark masses are important for tests of the standard model at the present
level of precision and have been already discussed in various areas of
particle phenomenology~\cite{PivovarovZ,chetm,larinm}. 

\subsection{Other types of induced currents}
Expressions for the induced currents with quantum numbers other than that of
the electromagnetic current $J^{PC}=1^{--}$ can be obtained in a similar way.
The axial current of fermions naturally appears in the standard model as a
result of an axial-vector interaction of fermions with $W$ and $Z$ bosons. The
corresponding induced current can be obtained as a derivative of the
respective effective action with respect to the $Z$ boson field. The leading
order diagram, however, contains only two external legs and is UV divergent.
In case of massless fermions this diagram leads to the anomalous
non-conservation of the axial current that also requires a strict definition of
the corresponding operator $\bar\psi\gamma_\mu\gamma_5\psi$ within perturbation
theory because its renormalization is not dictated by the Ward identities any
more. The leading corrections to the anomaly of the axial current for massless
fermions due to strong interactions were considered in Refs.~\cite{gabad,bos}
where implications of the Adler-Bardeen no-go theorem (non-renormalizability)
were discussed also for dimensional regularization and different definitions
of the axial current at the tree level. Higher order corrections were analyzed
in Ref.~\cite{laranom}. Explicit high order corrections to the expression for
the anomaly depend on the renormalization prescription for the composite
operator $\bar\psi\gamma_\mu\gamma_5\psi$ within perturbation theory.

The scalar current $\bar\psi\psi$ appears in an interaction vertex for the
Higgs boson and was intensively studied. The decay $H\to\gamma\gamma$ is
described by the effective interaction
\begin{equation}
\Delta L_H=g_S\alpha HFF
\end{equation}
with the parameter $g_S$ (remark the different to $g_s$, $S$ stands for the
scalar current) given by the corresponding one-loop diagram with two external
photons. Here $H$ is an interpolating field for the Higgs boson. The decay of
the Higgs boson into two photons is calculated up to high orders of
perturbation theory and mass expansion~\cite{hgg}. The application of the
effective potential technique to the analysis of the correlators of the scalar
gluonic currents which emerge in the decay of the Higgs boson into hadrons was
considered in Ref.~\cite{grav}.

\subsection{Results for the induced tensor current}
In the context of the effective action we consider a tensor current of the
form
\begin{equation}
J^{\mu\nu}=\bar\psi\sigma^{\mu\nu}\psi,\qquad
  \sigma^{\mu\nu}=\frac{i}2[\gamma^\mu,\gamma^\nu]
\end{equation}
and calculate its low-energy limit induced by a heavy quark loop. The
properties of this current are rather similar to those of the electromagnetic
current. Note that the classical vector mesons ($\rho$, $\omega$, $\phi$)
interact with this current and can be produced by it. We introduce an
interaction
\begin{equation}\label{lagten}
\Delta L_T=g_T\bar\psi\sigma^{\mu\nu}\psi F_{\mu\nu}
\end{equation}
in the Lagrangian of heavy quarks and readily find the effective action for
gauge fields induced by such a vertex. The low-energy limit at the one-loop
order reads
\begin{equation}
\Gamma_T=\frac{-g_Tg_s^3d_{abc}}{180m^3(4\pi)^2}
  \left(32\tr(FG^aG^bG^c)-15\tr(FG^a)\tr(G^bG^c)\right)
\end{equation}
with the same notations as in Eq.~(\ref{actqcd}).
According to the form of the effective interaction in Eq.~(\ref{lagten})
the induced current $J_{\mu\nu}$ is given by a derivative 
\begin{equation}
g_TJ^{\mu\nu}=-\frac{\delta\Gamma_T}{\delta F_{\mu\nu}}
\end{equation}
and explicitly reads
\begin{equation}\label{indten}
J^{\mu\nu}=\frac{-g_s^3d_{abc}}{180m^3(4\pi)^2}
  \left(32(G^aG^bG^c)^{\mu\nu}-15(G^a)^{\mu\nu}\tr(G^bG^c)\right).
\end{equation}
Note the lower power of the heavy quark mass in Eq.~(\ref{indten})
as compared to Eq.~(\ref{indvec}).

\section{Phenomenological applications}
High precision tests of the standard model remain one of the main topics of
particle phenomenology~\cite{PDG}. The recent observation of a possible signal
from the Higgs boson may complete the experimentally confirmed list of the
standard model particles~\cite{Higgs}. Because experimental data are becoming
more and more accurate, the determination of numerical values of the
parameters of the standard model Lagrangian will require more accurate
theoretical formulae. Recently an essential development in high-order
perturbation theory calculations has been observed. A remarkable progress has
been made in the heavy quark physics where a number of new physical effects
have been described theoretically with high precision. The cross section of
top--antitop production near the threshold has been calculated at the
next-to-next-to-leading order of an expansion in the strong coupling constant
and velocity of a heavy quark with an exact account for Coulomb interaction
(as a review, see Refs.~\cite{Hoang:2000yr,ppreview}). This theoretical
breakthrough allows for the best determination of a numerical value of the top
quark mass from experimental data. The method of Coulomb resummation resides
on a nonrelativistic approximation for the Green function of the
quark-antiquark system near the threshold and has been successfully used for
the heavy quark mass determination within sum rule
techniques~\cite{vol,leut,volnew}. Being applied to quarkonium systems this
method is considered to give the best estimates of heavy quark mass
parameters~\cite{PeninX,melye,hoamom,sumino}. Technically an enhancement of
near-threshold contributions to sum rules is achieved by considering integrals
of the spectral density of the heavy quark production with weight functions
which suppress the high-energy tail of the spectrum. The integrals with weight
functions $1/s^n$ for different positive integer $n$, $s=E^2$, where $E$ is
the total energy of the quark-antiquark system, are called moments of the
spectral density and are most often used in the sum rule
analysis~\cite{largen}. The interest in the precision determination of the
$c$-quark mass is especially high because this parameter introduces the
largest uncertainty to the theoretical calculation of the running
electromagnetic coupling constant at $M_Z$ which is one of the key quantities
for the constraints to the Higgs boson mass~\cite{runpiv}. We stress that in
the corresponding determination of the running electromagnetic coupling
constant at $M_Z$ based on direct integration of experimental data over the
threshold region, the sensitivity of the results to the $c$-quark mass is much
weaker~\cite{kunjalpha}.

For phenomenological applications of our results one therefore has to compute
the two-point correlation functions of induced currents. In the following we
show that there is a strong constraint on the order $n$ of the moment that can
be used in heavy quark sum rules. Because of the contribution of low-energy
gluons, only the first few moments exist if the theoretical expressions for
the correlators include the $O(\alpha_s^3)$ order of perturbation theory.

\subsection{The spectrum for the induced vector current correlator}
First we discuss the case of the vector current where the data are obtained
from $e^+e^-$ annihilation experiments and are rather precise. The basic
quantity for the analysis of a vector current $j^\mu=\bar q\gamma^\mu q$ of a
heavy fermion $q$ within sum rules is the vacuum polarization function
\begin{equation}\label{corvec}
12\pi^2i\int\langle Tj_\mu(x)j_\nu(0)\rangle e^{iqx}d^4x
  =(q_\mu q_\nu-g_{\mu\nu}q^2)\Pi(q^2).
\end{equation}
With the spectral density $\rho(s)$ defined by the relation 
\begin{equation}\label{spevec}
\rho(s)=\frac1{2\pi i}(\Pi(s+i0)-\Pi(s-i0)),\quad s>0,
\end{equation}
the dispersion representation
\begin{equation}\label{disrel}
\Pi(q^2)=\int\frac{\rho(s)ds}{s-q^2}
\end{equation}
holds. A necessary regularization and subtraction is assumed in
Eq.~(\ref{disrel}). The normalization of the vacuum polarization function 
$\Pi(q^2)$ in Eq.~(\ref{corvec}) is chosen such that one obtains the
high-energy limit $\lim_{s\to\infty}\rho(s)=1$ for a lepton. For the quark in
the fundamental representation of the gauge group $SU(N_c)$ the high-energy
limit of the spectral density reads $\rho(\infty)=N_c$. The integral in
Eq.~(\ref{disrel}) runs over the whole spectrum of the correlator in
Eq.~(\ref{corvec}) or over the whole support of the spectral density $\rho(s)$
in Eq.~(\ref{spevec}).

\begin{figure}\begin{center}\leavevmode
\put(0,0){(a)}
\epsfig{figure=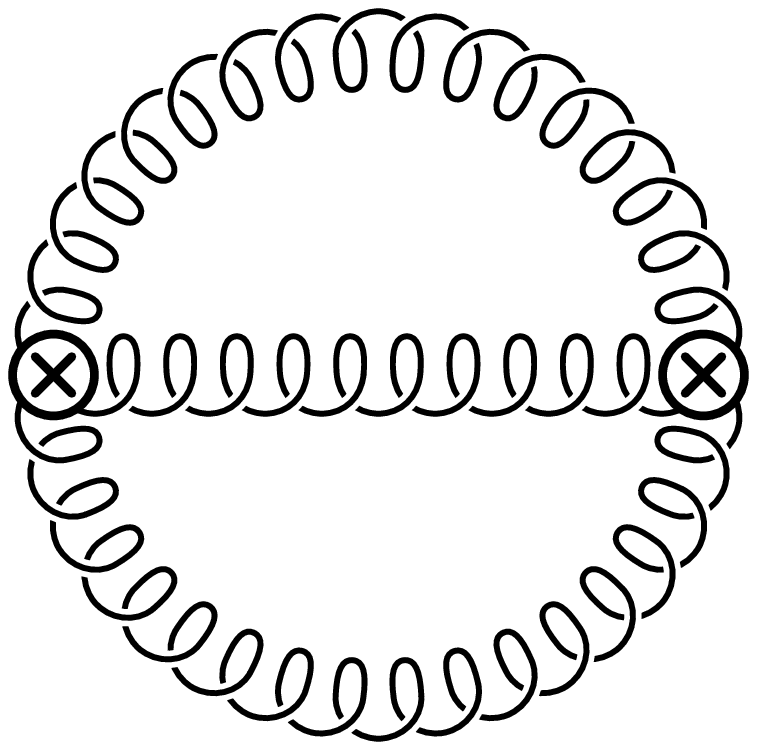, scale=0.4}\kern36pt
\put(0,0){(b)}
\epsfig{figure=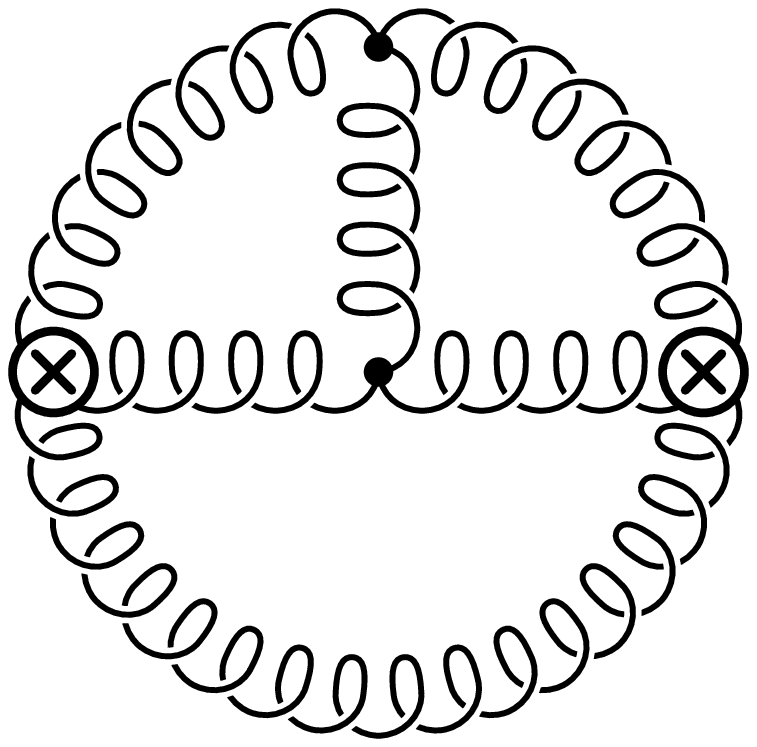, scale=0.4}
\caption{\label{fig2}Induced massless correlator diagrams}\end{center}
\end{figure}

A correlator of the induced vector current $J^\mu$ has the general form
\begin{equation}\label{glucor}
\langle TJ^\mu(x)J^\nu(0)\rangle=-\partial_\alpha\partial_\beta 
\langle T{\cal O}^{\mu\alpha}(x){\cal O}^{\nu\beta}\rangle
\end{equation}
where an explicit expression of the current as a derivative of the
antisymmetric operator ${\cal O}^{\mu\nu}$ has been employed. The resulting
correlator $\langle T{\cal O}^{\mu\alpha}(x){\cal O}^{\nu\beta}\rangle$ in
Eq.~(\ref{glucor}) contains only gluonic operators. Such correlators were
considered previously in the framework of perturbation
theory~\cite{PivovarovU,Okada}. In leading order of perturbation theory the
correlator in Eq.~(\ref{glucor}) has a topological structure of a sunset
diagram, as it is shown in Fig.~\ref{fig2}(a).
Technically, a convenient procedure of computing the sunset-type
diagrams is to work in configuration space~\cite{GrooteV}. We find 
\begin{equation}\label{xspace}
\langle TJ_\mu(x)J_\nu(0)\rangle
  =\frac{-34d_{abc}d_{abc}}{2025\pi^4m^8}\pfrac{\alpha_s}\pi^3
  \left(\partial_\mu\partial_\nu-g_{\mu\nu}\partial^2\right)\frac1{x^{12}}.
\end{equation}
A Fourier transform of the correlator in Eq.~(\ref{xspace}) gives the vacuum
polarization function in momentum space which reads
\begin{equation}\label{fourier}
12\pi^2i\int\langle TJ_\mu(x)J_\nu(0)\rangle e^{iqx}d^4x
  =(q_\mu q_\nu-g_{\mu\nu}q^2)\Pi(q^2)
\end{equation}
where at small $q^2$ ($q^2\ll m^2$)
\begin{equation}\label{piqvec}
\Pi(q^2)|_{q^2\approx 0}=C_g\pfrac{q^2}{4m^2}^4\ln\pfrac{\mu^2}{-q^2},\qquad
C_g=\frac{17d_{abc}d_{abc}}{243000}\pfrac{\alpha_s}\pi^3.
\end{equation}
For QCD with the colour group $SU(3)$ one has $d_{abc}d_{abc}=40/3$. The
spectral density of the vacuum polarization function $\Pi(q^2)$ in
Eq.~(\ref{fourier}) is given at small values for $s$ by
\begin{equation}\label{rhovec}
\rho(s)|_{s\approx 0}=C_g\pfrac{s}{4m^2}^4.
\end{equation}
Note that the spectral density in Eq.~(\ref{rhovec}) can be
found without an explicit calculation of its Fourier transform. Instead one
can use a spectral decomposition (dispersion representation) in configuration
space which was heavily employed for the analysis of sunset diagrams in
Ref.~\cite{GrooteV}. In this particular instance the spectral representation
of the correlator in configuration space reads
\begin{equation}
\frac{i}{x^{12}}=\frac{\pi^2}{2^8\Gamma(6)\Gamma(5)}\int_0^\infty s^4D(x^2,s)ds
\end{equation}
with $D(x^2,s)$ being the propagator of a scalar particle of mass $\sqrt s$,
\begin{equation}
D(x^2,m^2)=\frac{im\sqrt{-x^2}K_1(m\sqrt{-x^2})}{4\pi^2(-x^2)}
\end{equation}
where $K_1(z)$ is the McDonald function (a modified Bessel function of the
third kind, see e.g.\ Ref.~\cite{mac}). $\Gamma(z)$ is Euler's gamma function.

An asymptotic behaviour of the spectral density of the corresponding
contribution for large energies (where the limit of massless quarks can be
used) enters the expression for the ratio $R(s)$ of $e^+e^-$ annihilation into
hadrons and has been known since long ago~\cite{kataev,chet,surg}. This term
is usually called light-by-light (lbl) contribution and reads
\begin{equation}\label{asym}
R^{\rm lbl}(s)=\pfrac{\alpha_s}\pi^3\frac{d_{abc}d_{abc}}{1024}
  \left(\frac{176}3-128\zeta(3)\right).
\end{equation}
Here $\zeta(z)$ is the Riemann $\zeta$ function with $\zeta(3)=1.20206\ldots$
The contribution to the spectral density given in Eq.~(\ref{asym}) is negative
while our result given in Eq.~(\ref{rhovec}) is positive as it should be the
case for the spectral density of the electromagnetic current as a Hermitean
operator.

\subsection{The spectrum for the induced tensor current correlator}
The results for the correlator of the tensor current given in
Eq.~(\ref{indten}) are slightly more complicated. The correlator reads
\begin{eqnarray}\label{corten}
\lefteqn{12\pi^2i\int\langle Tj_{\mu\nu}(x)
  j_{\alpha\beta}(0)\rangle e^{iqx}d^4x\ =\ (g_{\mu\alpha}g_{\nu\beta}
  -g_{\mu\beta}g_{\nu\alpha})\Pi_g(q^2)}\nonumber\\&&
  +(g_{\mu\alpha}q_\nu q_\beta-g_{\mu\beta}q_\nu q_\alpha
  -g_{\nu\alpha}q_\mu q_\beta+g_{\nu\beta}q_\mu q_\alpha)\Pi_q(q^2)
\end{eqnarray}
where two scalar amplitudes are possible now. One finds
\begin{equation}\label{piqten}
\Pi_g(q^2)=-\frac{q^2}2\Pi_q(q^2)=\frac{61d_{abc}d_{abc}}{729000}
  \pfrac{\alpha_s}\pi^3\frac{-q^2}4\pfrac{q^2}{4m^2}^3\ln\pfrac{\mu^2}{-q^2}.
\end{equation}
The physical content of the amplitudes $\Pi_g(q^2)$ and $\Pi_q(q^2)$ is
related to contributions of the states with $J^{PC}=1^{--}$ and
$J^{PC}=1^{+-}$, resp. Note that the sum rule analysis for the mesons with
quantum numbers $J^{PC}=1^{+-}$ has been done in Ref.~\cite{ovch} with quark
interpolating currents. From the present results we also see a possibility to
use gluonic currents as interpolating operators for such mesons. The validity
of such a description depends strongly on the strength of the interaction of
the meson in question with the corresponding interpolating operator which is
difficult to estimate independently.

Note that there are only two independent gluonic operators to construct the
induced currents under consideration. The electromagnetic current is given by
a derivative of a special linear combination of these operators while the
tensor current is given by a linear combination of operators themselves. There
is one more current relevant to the situation. It originates from the Gordon
decomposition of the electromagnetic current (see e.g.\ Ref.~\cite{Itzykson})
\begin{equation}\label{gordon}
2m\bar\psi\gamma^\mu\psi=\partial_\nu(\bar\psi\sigma^{\mu\nu}\psi) 
  +\bar\psi\ i\!\!\buildrel\leftrightarrow\over D\!{}^\mu\psi.
\end{equation}
This relation retains for the induced currents as well. The left hand side and
the right hand side of Eq.~(\ref{gordon}) have different parity as for the
number of Dirac $\gamma$-matrices between spinor fields which is reflected in
an additional factor $m$ at the left hand side of Eq.~(\ref{gordon}). In the
massless limit these types of currents are alien and can never mix. At the
level of induced currents the Dirac structure of the initial heavy quark
currents is reflected in different degrees of suppression by the heavy quark
mass $m$.

\subsection{The spectrum for a mixed current correlator}
Having both currents at hand, one can study a mixed correlator of the form 
\begin{equation}\label{cormix}
12\pi^2i\int\langle TJ_{\mu}(x)J_{\alpha\beta}(0)\rangle e^{iqx}d^4x
  =i(g_{\mu\alpha}q_\beta-g_{\mu\beta}q_\alpha)\Pi_M(q^2)
\end{equation}
with a single scalar amplitude $\Pi_M(q^2)$. Such mixed correlators are useful
in sum rule applications~\cite{misuse}. One finds
\begin{equation}\label{piqmix}
\Pi_M(q^2)=\frac{67d_{abc}d_{abc}}{972000}\pfrac{\alpha_s}\pi^3
  \frac{-q^2}{4m}\pfrac{q^2}{4m^2}^3\ln\pfrac{\mu^2}{-q^2}
\end{equation}
The physical content of the amplitude is given by the $J^{PC}=1^{--}$
resonance, i.e.\ by the $\Upsilon$-family in case of $b$ quarks.

\section{Moments of the spectral density}
As mentioned earlier, the moments of the spectral density $\rho(s)$ of the
form
\begin{equation}\label{mom}
{\cal M}_n=\int\frac{\rho(s)ds}{s^{n+1}}
\end{equation}
are usually studied within the sum rule method for heavy quarks~\cite{largen}.
These moments are related to the derivatives of the vacuum polarization
function $\Pi(q^2)$ at the origin,
\begin{equation}\label{momder}
{\cal M}_n=\frac1{n!}\pfrac{d}{dq^2}^n\Pi(q^2)\bigg|_{q^2=0}.
\end{equation}
Such moments are chosen in order to suppress the high energy part of the
spectral density $\rho(s)$ which is not measured accurately in the experiment.
Within the sum rule method one assumes that the moments in Eq.~(\ref{mom}) can
be calculated for any $n$ or, equivalently, that the derivatives in
Eq.~(\ref{momder}) exist for any $n$. The existence of moments seems to be
obvious because one implicitly assumes that the spectral density $\rho(s)$ of
the heavy quark electromagnetic currents vanishes below the two-particle
threshold $s=4m^2$ which means that the vacuum polarization function
$\Pi(q^2)$ of heavy quarks is analytic in the whole complex plane of $q^2$
except for the cut along the positive real axis starting from $4m^2$. This
assumption about the analytic properties of the vacuum polarization function
$\Pi(q^2)$ is known to be wrong if a resummation of Coulomb effects to all
orders of perturbation theory is performed: as a result of such a resummation
the Coulomb bound states appear below the perturbation theory threshold
$s=4m^2$.

\subsection{Infrared singular behaviour of the moments}
The qualitatively new feature of effective currents given in
Eqs.~(\ref{indvec}) and~(\ref{indten}) is that they are expressed through
massless fields. Therefore the spectrum of the two-point correlators of these
currents start at zero energy. This feature drastically changes the analytic
structure of the two-point correlators of these currents and, in particular,
their infrared (IR) or small $s$ behaviour because of the branching point
(cut) singularity of $\Pi(q^2)$ at the origin $q^2=0$. This new feature of
having a nonvanishing spectrum below the formal tree-level two-particle
threshold which appears at the $O(\alpha_s^3)$ order of perturbation theory for
induced current correlators has important phenomenological consequences.
Indeed, such a change of the analytic structure of induced current correlators
affects strongly the theoretical expressions for some observables usually
employed in heavy quark physics for the precision determination of the
parameters of heavy quarks and their interactions.

Because of the low-energy gluon contributions, the large $n$ moments of the
spectral density in Eq.~(\ref{mom}) do not exist and cannot be used for
phenomenological analyses. Caused by the factor $(s/4m^2)^4$ in
Eq.~(\ref{mom}) the moments become IR singular for $n\ge 4$ in case of the
induced vector current. This can already be seen by looking at the factor
$1/m^4$ in the induced vector current in Eq.~(\ref{indvec}). For the induced
tensor current the corresponding moments start to diverge earlier because of a
weaker suppression by the heavy quark mass, the corresponding factor in
Eq.~(\ref{indten}) is $1/m^3$ instead of $1/m^4$. Therefore, in this case the
moments become IR singular already for $n\ge 3$. Note that in early
considerations of sum rules quite large $n$ were used. For instance, the
numerical value of the gluon condensate was extracted from sum rules for the
moments with $n\sim 10\div 20$~\cite{largen,rad}. In view of our result on
the low-energy behaviour of the spectral density, one has either to limit the
accuracy of theoretical calculations for the moments to the $O(\alpha_s^2)$
order of perturbation theory which seems insufficient for a high precision
analysis of quarkonium systems (especially if the Coulomb resummation to all
orders is performed) or to use only a few first moments. For small $n$,
however, the high-energy contribution, which is not known experimentally with
a reasonable precision, is not sufficiently suppressed and introduces a large
quantitative uncertainty into the sum rules for the moments. An analysis based
on finite energy sum rules is free from such a problem and can be used in
phenomenological applications~\cite{KrasnikovZ}.

Note in passing that there is no low-energy gluon contribution (and therefore
no low-energy divergence problem) for correlators of the currents containing
only one heavy quark with mass $m$. The spectrum of such correlators starts at
$m^2$ and there are no massless intermediate states contributing to the
correlator in perturbation theory (see e.g.\ Ref.~\cite{bar}). The theoretical
expressions for such correlators can be used for high precision tests of
theoretical predictions when the accuracy of experimental data in
corresponding channels will improve in the future.

\subsection{An infrared safe moment definition}
The infinite-integration sum rules with large $n$ can be retained at high
orders of perturbation theory if an appropriate cutoff at small energies is
introduced. This can be readily achieved by calculating the moments (or
derivatives) at some Euclidean point $q^2=-\Delta$~\cite{reinders}. Indeed,
for the regularized moments 
\begin{equation}\label{momcut}
{\cal M}_n(\Delta)=\frac1{n!}\pfrac{d}{dq^2}^n\Pi(q^2)\bigg|_{q^2=-\Delta}
  =\int\frac{\rho(s)ds}{(s+\Delta)^{n+1}}
\end{equation}
there is no divergence at small $s$. However, the regularization parameter
$\Delta$ cannot be arbitrary small. The reason is that the resulting
correlator of gluonic currents in Eq.~(\ref{glucor}) is essentially
normalized at $\mu^2=\Delta$ when radiative corrections are taken into
account, like the diagram shown in Fig.~\ref{fig2}(b). The corrections are
known to be large and, therefore, $\Delta$  is much larger than the expected
IR scale in quark channels~\cite{PivovarovU}. This observation makes the
phenomenological analysis based on the regularized sum rules in
Eq.~(\ref{momcut}) unprecise even for reasonably large $n$ because the
continuum contribution to moments is not suppressed for large values of
$\Delta$. The suppression of the high-energy tail can be enhanced by
constructing a more special kind of moments which exploit the explicit
behaviour of the spectral density at low energy. The IR safe moments with a
more efficient suppression of high-energy tail for the induced vector current
have the form
\begin{equation}\label{mommod}
{\tilde {\cal M}}_n(\Delta)=\int\frac{\rho(s)ds}{s^3(s+\Delta)^{n+1}}
\end{equation}
It seems that even in such an optimized form these moments cannot be used for
the $c\bar c$-system, i.e.\ for analyzing $J/\psi$ resonances because the
suppression is insufficient for achieving the precision goals necessary for
the $c$-quark mass determination. For the $b\bar b$-system, i.e.\ for the
analysis of $\Upsilon$-resonances the dependence on the cutoff $\Delta$ is
numerically essential for moments at large $n$ and for $\Delta$ as large as
$\Delta=1\div 2\GeV^2$ which is still small according to the estimates of
radiative corrections in gluonic channels. We would like to remind the reader
that Coulomb poles which are essential for the analysis of the
$\Upsilon$-resonance and $b\bar b$-production near threshold give
contributions which are formally of the $O(\alpha_s^3)$ order (for the value
of the Coulombic wave function at the origin see e.g.\ Ref.~\cite{PeninX}),
an order which coincides with the order of corrections considered here (see
Eq.~(\ref{piqvec})). The theoretical expressions for the correlators in the
scalar channel where below threshold corrections start at the $O(\alpha_s^2)$
order are more sensitive to these special below threshold contributions.
However, data in the scalar channel are considerably worse than those in the
vector channel and there is no possibility of a high precision analysis on the
scalar channel at present.

\section{Conclusions}
We have presented corrections to the heavy quark currents induced by a virtual
heavy quark loop and expressed through the gluon operators. We have considered
the electromagnetic current and the tensor current closely related to it.
Heavy quark loop induced corrections first appear at $O(\alpha_s^3)$ order of
perturbation theory and are given by the $1/m^4$ term of the mass expansion in
case of the electromagnetic current and by the $1/m^3$ term in case of the
tensor current. The spectra of the correlators of such induced currents start
at zero energy. This fact makes impossible the standard analysis of the moment
sum rules at $O(\alpha_s^3)$ order of perturbation theory for $n\ge 4$ in case
of the electromagnetic current and for $n\ge 3$ in case of the tensor current. 

\subsection*{Acknowledgements}
The work is partially supported by the Russian Fund for Basic Research under
contracts 99-01-00091 and 01-02-16171. A.A.~Pivovarov is an Alexander von
Humboldt fellow. S.~Groote acknowledges a grant given by the Deutsche
Forschungsgemeinschaft, Germany.

\end{document}